  \documentclass[prl,aps,twocolumn,showpacs]{revtex4}
\usepackage[dvips]{graphicx}
\usepackage{dcolumn}%
\usepackage{amsmath}%
\usepackage{amsfonts}%
\usepackage{amssymb}
 \usepackage[usenames]{color}
\providecommand{\U}[1]{\protect\rule{.1in}{.1in}}
\newcommand{\be}{\begin{equation}}
\newcommand{\ee}{\end{equation}}
\newcommand{\bea}{\begin{eqnarray}}
\newcommand{\eea}{\end{eqnarray}}
\newcommand{\bt} {\begin{tabular}}
\newcommand{\et} {\end{tabular}}
\newcommand{\nn}{ \nonumber}
\newcommand{\ds}{\displaystyle}
\newcommand{\ba} {\begin{array}}
\newcommand{\ea} {\end{array}}
\begin{document}

\title{Thermoelectric efficiency of single-molecule junctions with long molecular linkers}

\author{  Natalya A. Zimbovskaya}

\affiliation
{Department of  Physics and Electronics, University of Puerto Rico-Humacao, CUH Station, Humacao, PR 00791, USA}

\begin{abstract}
We report results of theoretical studies of thermoelectric efficiency of single-molecule junctions with long molecular linkers. The linker is simulated by a chain of identical sites described using a tight-binding model. It is shown that thermoelectric figure of merit ZT strongly depends on the bridge length,  being controlled by the lineshape of electron transmission function within the tunnel energy range  corresponding to HOMO/LUMO transport channel. Using the adopted  model we demonstrate that ZT may significantly increase as the linker lengthens, and that gateway states on the bridge (if any) may noticeably affect the length-dependent ZT. Temperature dependences of ZT for various bridge lengths are analyzed. It  is shown that broad minima emerge in ZT versus temperature curves whose positions are controlled by the bridge lengths.
\end{abstract}

\date{\today}
\maketitle


\section{I. Introduction}

Significant effort which has been applied to study various aspects of electron transport through molecular systems mostly stems from possible applications of these systems in molecular electronics.  Molecular electronics provides a general platform to manufacture atomic-scale energy conversion devices operating by means of control of charge and thermal currents through molecules and other similar systems including carbon-based nanostructures and quantum dots \cite{1,2,3,4,5}.
       A single-molecule junction consists of a couple of metallic/semiconducting electrodes linked by a molecular bridge. As known, single-molecule junctions hold promise for enhanced efficiency of heat-to-electric energy conversion. Therefore,  thermoelectric properties of these systems have been explored both theoretically and experimentally \cite{6,7,8}. Numerous works were focused on Seebeck effect which is directly responsible for conversion of heat to electric energy \cite{9}. The effect occurs when a thermal gradient is applied across a system inducing a charge current. In molecular junctions, Seebeck effect is measured by recording the voltage $ \Delta V $ which stops the thermally induced current \cite{10,11}. When the temperature difference  $ \Delta T $ between the ends of the system is sufficiently small, the efficiency of energy conversion is characterized by the thermopower $ S $ which is equal the the ratio $\ds  - \frac{\Delta V}{ \Delta T} $ and by a  dimensionless thermoelectric figure of merit $ ZT: $
\be
ZT = \frac{S^2 GT}{\kappa}.  \label{1}
\ee
Here, $ G $ is the electronic electrical conductance, $ T $ is the average temperature of the considered system and $ \kappa $ is the thermal conductance which generally includes contributions from both electrons and phonons $(\kappa_{el} $ and $\kappa_{ph},$ respectively). 
It was shown that thermoelectric properties of single-molecule junctions may be affected by dephasing/inelastic effects \cite{12,13,14}, by the molecular bridge geometry \cite{15,16}, by Coulomb interactions between electrons on the bridge \cite{17,18,19,20,21,22}, by molecular vibrations \cite{23,24,25,26}, and by quantum interference \cite{27,34}. 

   Among thermoelectric properties of single-molecule junctions one may separate out the thermoelectric figure of merit ZT which is directly responsible for the thermoelectric efficiency of the system \cite{32}.  This important thermoelectric characteristic was previously studied in several works (see  e.g.  Refs. \cite{16,17,19,33,34,35}. However, these studies were focused on the effects of  conformational changes of the bridge geometry \cite{16}, electron-electron interactions \cite{17,19,33,35} and/or quantum interference \cite{34}. The effect of the bridge length on ZT was not thoroughly investigated so far. It is established that
	in systems  where the molecular linker is a chain-like structure consisting of several identical units  (the units being e.g. benzene or phenyl rings),  Seebeck coefficient strongly depends on the linker length \cite{10,11,16,28,29,30,31}. Therefore,  one may expect the figure of merit to be length-dependent as well. The purpose of the present work is to explore the relationship between the thermoelectric efficiency of a single-molecule junction characterized by $ ZT $ and the length of the molecular  bridge.

\section{II. Model and results} 

In the following analysis we adopt a simple model for the molecular bridge. We simulate it by a periodical chain including $ N $ sites, the  number of sites indicating the bridge length. For each site we assign an on-site energy $ E_i. $ We assume that each site (except  terminal ones) is coupled to its nearest neighbors with the coupling strengths $ \beta_{i-1,i} $ and $ \beta_{i,i+1}\ (2 \leq i \leq N -1).$ To further simplify the model we put $ E_i = E_0 $ and $ \beta_{i-1,i} = \beta_{i,i+1} = \beta $ for all relevant sites in the chain. The terminal sites are coupled to the electrodes through imaginary self-energy terms $\ds - \frac{i}{2} \Gamma_{L,R}.$ Within the  wide band approximation $ \Gamma_{L,R} $ do not depend on the tunnel energy $E. $ Below we consider symmetrically coupled systems, so we put  $ \Gamma_L = \Gamma_R = \Gamma. $  Similar models are often used to mimic molecular bridges consisting of repeating units provided that $ \pi$--$\pi $ coupling dominates electron transport \cite{36}. Then the parameter $ \beta $ characterizes the coupling between adjacent $ \pi $ orbitals. 
 The schematics of the adopted model showing all relevant parameters is presented in Fig. 1. Such models were used to represent realistic metal-molecule junctions such as those made out of  gold electrodes linked by alcane/oligophenyl chains \cite{31} or by CSW-470-bipyridine molecules \cite{16}. The model was previously employed to analyze length-dependent thermopower \cite{16,31,37}  and thermally induced charge current \cite{38} in single-molecule junctions.

 Also, we assume coherent electron tunneling to be a predominant transport mechanism. We omit from consideration the effect of molecular vibrations and the phonon contribution to the thermal conductance. The smallness of $ \kappa_{ph} $ in single-molecule junctions may be attributed to the fact that  a significant mismatch between phonon density of states on the electrodes and phonon modes associated with molecular vibrations exists in many such systems  \cite{19,39}.  Therefore, we may  expect $ \kappa_{ph} $ to be small when the temperature of the system takes on values of the order of, or lower than room temperature. 

\begin{figure}[t] 
\begin{center}
\includegraphics[width=8cm,height=8.5cm]{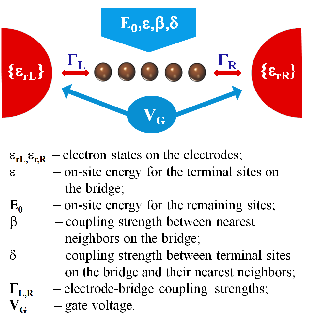}
\caption{ Schematics of the metal-molecule-metal junction used to analyze the length-dependent thermoelectric figure of merit. Indicated parameters represent relevant energies. 
}
 \label{rateI}
\end{center}\end{figure}

Within the accepted model, the electron transmission $\ds \tau(E) = \frac{\Gamma^2}{4} \big|G_{1N}(E)\big|^2 $ where $ G_{1N} $ is the corresponding matrix element of the retarded Green's function for the molecular bridge:
\be
G = (E - H - i \Gamma)^{-1}   \label{2}
\ee
We remark that matrix elements of the bridge Hamiltonian $ H $ are often computed using electron structure calculations (see e.g.  Refs. \cite{15,28,29,30,31}. Nevertheless, tight-binding and Lorentzian models were successfully used to explain length dependences of the thermopower in single-molecule junctions observed in experiments \cite{28,29,30,31,33,34}. This gives grounds to believe that the presently employed model may lead to useful qualitative results in analyzing the behavior of $ ZT $ in these systems.

Within the accepted tight-binding model the Hamiltonian $ H $ is represented by $ N \times N $ matrix:
\be 
H =  \left [\ba{cccccc}
\ds E_0 -  \frac{i\Gamma}{2} &  \beta & 0 & 0 & \cdots & 0
 \\
\beta  &  E_0  &  \beta  &  0  &  \cdots  &  0
\\
0  & \beta  &  E_0  &  \beta  &  \cdots  &  0
\\
\cdots & \cdots & \cdots & \cdots & \cdots & \cdots
\\
0 & 0 & \cdots & \beta  &  E_0  &  \beta
\\
0 & 0 & \cdots &  \cdots   &  \beta  & \ds E_0 -  \frac{i\Gamma }{2}
\\ 
\ea \right ] .  \label{3}
\ee

Within the regime of linear response of the molecular junction to the applied thermal gradient $(\Delta T  \ll T), $ the main characteristics of thermoelectric transport are is given by \cite{40}:
\be
  G = e^2 L_0,   \label{4}
\ee
\be
  S = -\frac{1}{eT} \frac{L_1}{L_0},  \label{5}
	\ee
	\be
 \kappa_{el} = \frac{1}{T} \left(L_2 - \frac{L_1^2}{L_0} \right)  .  \label{6} 
\ee
Correspondingly:
\be
ZT = \frac{L_1^2}{L_0L_2 - L_1^2}   . \label{7}
\ee 
Here, the coefficients $ L_n $ have the form:
\be
L_n = - \frac{1}{h} \int (E - \mu)^n \tau(E) \frac{\partial f}{\partial E} dE.  \label{8}
\ee
In this expression, $ \mu $ is the chemical potential of electrodes, and $ f(E) $ is the Fermi distribution function computed at the temperature $ T $ and chemical potential $ \mu . $

\begin{figure}[t] 
\begin{center}
\includegraphics[width=8.3cm,height=7.4cm]{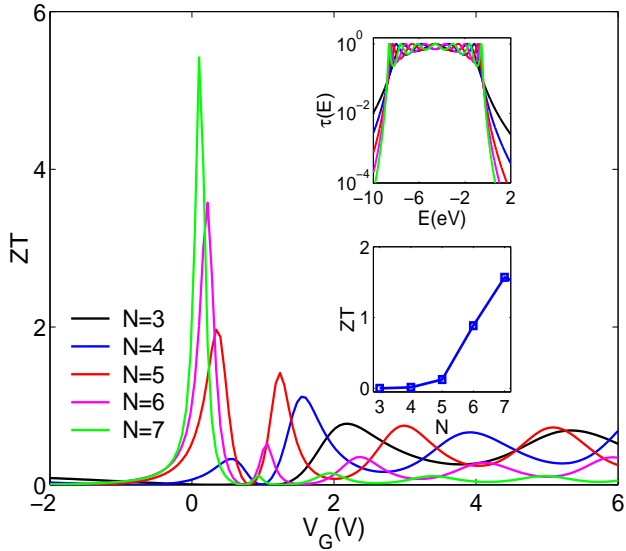} 
\caption{ Thermoelectric figure of merit of an unbiased single-molecule junction as a function of gate voltage $ V_G. $ The upper inset shows the transmission plotted as a function of the tunnel energy for several different bridge lengths.The lower inset displays  $ ZT $ as a function of the bridge length at  $ V_G = 0. $ All curves are plotted at $ T = 300 K,\  E_0 = - 4.6 eV,\ \beta = 2.2 eV, \ \Gamma = 2.5 eV. $ It is accepted that $ \mu = 0. $
}
 \label{rateI}
\end{center}\end{figure}

Thermally induced charge carriers travel through the system using the highest occupied molecular linker orbital (HOMO) or its lowest unoccupied orbital (LUMO) as transport channels.  For certainty, in further analysis we assume HOMO to be the transport channel, as illustrated by the transmission profile plotted in Fig. 2. One observes that HOMO is located slightly below the electrodes chemical potential $ \mu = 0. $
 When the energies $E_{H,L} $ associated with these orbitals noticeably differ from $ \mu \ (|E_{H,L} - \mu| > kT,\ k $ being the Boltzmann's constant) one can estimate coefficients $ L_n $ using Sommerfeld expansion \cite{41}:
\begin{align}
&   L_0 = \frac{1}{h} \tau(\mu),
\nn\\ &
L_1 = \frac{1}{h} \frac{(\pi kT)^2}{3} \frac{\partial \tau}{\partial E} \Big|_{E=\mu}, 
\nn\\ &
 L_2 = \frac{1}{h} \frac{(\pi kT)^2}{3} \tau (\mu).    \label{9}
\end{align}
 In the considered case, when the molecular bridge is simulated by a simple chain of identical sites, the transmission $ \tau (\mu) $ shows an exponential decrease as the  bridge length increases   provided that $\mu $ is located sufficiently far away from molecular resonances. 
 Near $ E = \mu, \ \tau(E) $ may be presented in the form: $ \tau(E) = A(E) \exp[- \eta(E) N] $ where $ A(E) $              determines length-independent contributions to the thermopower (see Eq. (\ref{10}) and to $ \ln G $ (see Eqs. (\ref{4}), (\ref{9}).  Both $ A (E) $  		and $ \eta(E) $ are slowly varying functions of energy.  This approximation follows from the theory of off-resonant tunneling, and it was confirmed by experiments on several long molecules \cite{29,42,43,44}. 
		The transmission behavior results in a similar exponential fall of both $ G $ and $ \kappa_{el}, $ whereas the thermopower $ S $ becomes a linear function of the bridge length \cite{29}:
\be
S = - \frac{(\pi kT)^2}{3eT} \left\{\frac{\partial \ln A}{\partial E} \Big|_{E = \mu} - \frac{\partial \eta}{\partial E} \Big|_{E =\mu} N \right\}.   \label{10}
\ee
Therefore, the dependence of  $ ZT $ on the bridge length is mostly determined by behavior of the factor $ S^2 $ in the Eq. (\ref{1}) and one may expect $ ZT $ to rather slowly enhance as the bridge lengthens. Computations carried out within the chosen model confirm this result as shown in Fig. 2.

\begin{figure}[t] 
\begin{center}
\includegraphics[width=8.3cm,height=7.4cm]{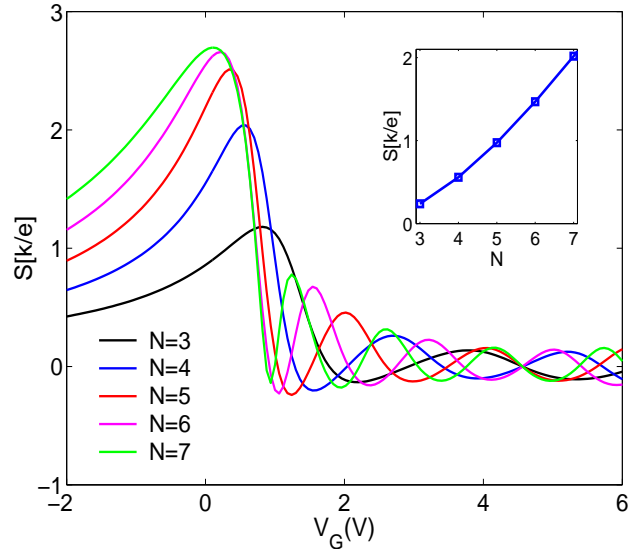} 
\caption{ Thermopower of an unbiased single-molecule junction as a function of the gate voltage. Inset shows the length-dependent thermopower plotted at $ V_G = 0. $ All relevant parameters used in plotting the curves have the same values as those in Fig. 2.
}
 \label{rateI}
\end{center}\end{figure}

A gate voltage applied to the system shifts molecular bridge energy levels with respect to the chemical potential $ \mu. $ At small positive $ V_G $ HOMO approaches $ \mu $ from below which brings a significant enhancement of the thermopower displayed in Fig. 3. Consequently, we observe a maximum in $ ZT $ near $ V_G  = 0 $ shown in Fig. 2. The peak height is distinctly dependent on the bridge length. The greater is $ N $ the sharper and  higher  the peak becomes. For  $ N \geq 5,$ maximum value of $ ZT $  significantly exceeds $ 1. $
        One observes that all computations carried on to plot curves shown in Fig. 2 (and in all subsequent figures) are performed assuming that $3 \leq N \leq 7.$ This assumption is made to focus on  realistic molecular bridge lengths close to those reported in experiments \cite{10,11,28,29,30,31}. The adopted model may be used to compute characteristics of thermoelectric transport for an arbitrary number of the bridge sites. However, it is difficult to make suggestions concerning the behavior of thermoelectric characteristics of single-molecule junctions with very long bridges in advance. Their behavior may vary depending on the relative positions of the relevant transmission peaks and the chemical potential of electrodes, among other factors. In some cases, ZT may saturate at large values on $ N $,  in other cases it may decrease when $N $ exceeds some particular  value. Future experiments may bring more light to this matter.

As $ V_G $ strengthens, transport channels associated with other molecular orbitals come into play. As one of these levels approaches $ \mu $ from below, it starts to serve as a transport channel for holes pushed through the system by the temperature gradient. Further increase in $ V_G $ forces the level to shift to a position above $ \mu $ (but still close to the latter) and to serve as a transport channel for electrons. When the level is shifted farther away from $ \mu,$ it ceases to participate in thermally-induced transport. As a result, one observes oscillations in the thermopower similar to those studied for multilevel qualtum dots \cite{45}. These oscillations are displayed in Fig. 3. Thermopower oscillations caused by alterations in the electrodes chemical potential of a single-molecule junction were observed in experiments \cite{46}. The oscillations of  thermopower lead to emergence of sequence of peaks in $ ZT $  (see Fig. 2).  

Linear relationship between the single-molecule junction thermopower and molecular bridge length such as given by Eq. (\ref{10}) is not the universal one. Distinctly nonlinear length dependences of thermopower were observed in experiments (see e.g.  Ref. \cite{31}). It was suggested that nonlinearity in the relationship between $ S $ and $ N $ appears due  to special properties of terminal sites on the bridge. Within a tight-binding model, we may account for these properties by assigning to these sites on-site energies which differ from those assigned to the remaining sites $(E_1 = E_N = \epsilon,\ E_i = E_0,\ 2 \leq i \leq N -1) $ and by assuming that their coupling with nearest neighbors is characterized by the coupling parameter $ \delta $ which is different from the parameter $ \beta $ describing coupling between other sites. Then the tight-binding Hamiltonian $ H $ accepts the form:
\be   
H =  \left [\ba{cccccc}
\ds \epsilon -  \frac{i\Gamma}{2} &  \delta & 0 & 0 & \cdots & 0
 \\
\delta  &  E_0  &  \beta  &  0  &  \cdots  &  0
\\
0  & \beta  &  E_0  &  \beta  &  \cdots  &  0
\\
\cdots & \cdots & \cdots & \cdots & \cdots & \cdots
\\
0 & 0 & \cdots & \beta  &  E_0  &  \delta
\\
0 & 0 & \cdots &  \cdots   &  \delta  &\ds  \epsilon -   \frac{i\Gamma }{2}
\\ 
\ea \right ] .  \label{11}
\ee

\begin{figure}[t] 
\begin{center}
\includegraphics[width=8.3cm,height=7.4cm]{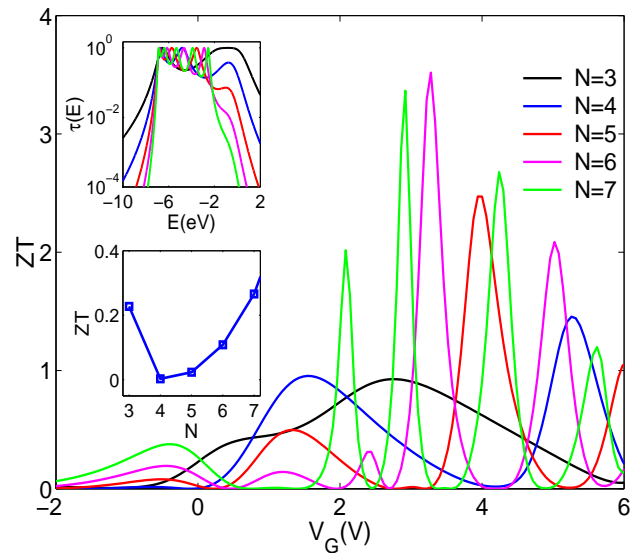} 
\caption{The effect of gateway states on the length-dependent thermoelectric figure of merit plotted as a function of the gate voltage. The upper inset shows the electron transmission presented as a function of tunnel energy. The lower inset displays $ ZT $ as a function of the bridge length at $ V_G = 0 .$ Curves are plotted assuming $ E_0 = - 4.47 eV,\ \epsilon = - 1.85 eV,\ \beta = 2.28 eV,\  \delta = 1.27 eV,\ \Gamma = 2.85 eV,\ T = 300 K. $  
}
 \label{rateI}
\end{center}\end{figure}

Using the bridge Hamiltonian modified in this manner to calculate the electron transmission, one observes that special characteristics of terminal sites may significantly affect the transmission profile. In the following analysis we use $ E_0 =  -4.47 eV,\ \epsilon = - 1.85 eV,\ \Gamma =2.85 eV,\ \beta = 2.28 eV $ and $ \delta = 1.27 eV. $ These are the values computed for single-molecule junctions with gold electrodes linked by molecular bridges consisting of sequences of oligophenyls terminated with trimethyltin end groups \cite{31}. The computations were carried out using density functional theory. Accepting these values for relevant energies, the HOMO lineshape becomes significantly distorted compared to that of a simple bridge chain (see inset in Fig. 4). 
We remark that in the chosen junction gateway states originate from Au-C bonds. However, other single-molecule junctions where the linker is coupled to the electrodes through more energetically favored bonds (e.g. Au-S) may be analyzed using the tight-binding Hamiltonian of the form (\ref{11}), and  similar distortion of HOMO/LUMO lineshape may be revealed.

As discussed in earlier works \cite{31,37,38}, the specific HOMO profile caused by terminal (gateway) sites on the bridge may affect length dependences of the thermopower and the thermally induced charge current. Also, the presence of gateway sites may significantly change $ ZT. $ As shown in Fig. 4, at $ V_G = 0,\ ZT $ takes on values which are  considerably smaller than those computed for a simple chain-like bridge with close values of relevant energies and displayed in Fig. 2. Nevertheless, at certain values of the gate voltage $ ZT $ shows  peaks of considerable height. Another effect of gateway states is a minimum which emerges in the length-dependent $ ZT $ at a certain bridge length. Such minimum does not appear in the case of a simple bridge chain.
 Changes in length-dependent thermoelectric characteristics (including the minimum in ZT) appear due to the changes in the relationship between $ \tau(\mu) $ and $\ds \frac{\partial \tau}{\partial E} \Big|_{E = \mu}$ which originate from the distortion of HOMO lineshapes.

\begin{figure}[t] 
\begin{center}
\includegraphics[width=8.3cm,height=7.3cm]{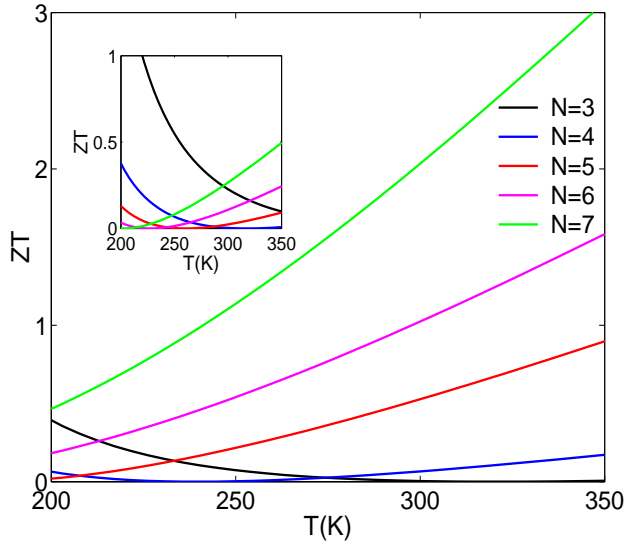} 
\caption{Temperature dependences of $ ZT $ for different bridge lengths. Curves are plotted assuming $V_G = 0,\ \mu =0,\ E_0 = -4.6eV,\ \beta = 2.2 eV,\ \Gamma = 2.5 eV. $ Inset shows the effect of gateway states on the temperature dependences of $ ZT. $ Curves in the inset are plotted ar $ V_G = 0,\  
 \mu = 0,\ E_0 = - 4.47 eV,\ \epsilon = - 1.85 eV,\ \beta = 2.28 eV,\ \delta = 1.27 eV,\ \Gamma = 2.85 eV. $
}
 \label{rateI}
\end{center}\end{figure}

As follows from approximations for coefficients $ L_n, $ given by Eq. (\ref{9}), at low temperatures the figure of merit increases as the temperature rises, $ ZT $ being proportional to $ (kT)^2 . $ At higher temperatures, next terms in Sommerfeld expansions for $ L_n $ must be taken into account.  These higher order in $ kT $  terms are proportional to higher order derivatives of $ \tau(E) $ taken at $ E = \mu. $ Therefore, the profile of electron transmission function near $ E = \mu $ may significantly affect temperature dependences of $ ZT. $ In the present work, we have chosen all relevant energies in such a way that all energy levels of the bridge are situated below $ E = \mu $ (see insets in Figs. 2, 3). This asymmetry was deliberately introduced to emphasize the variations in the transmission profile occurring due to variations in the bridge length as well as those caused by gateway states. As a result, the transmission rapidly falls within the relevant energy range, and this behavior is responsible for broad minima emerging in $ ZT $ temperature dependences at certain temperatures  shown in Fig. 5. So, at moderately high temperatures, the thermoelectric efficiency of the considered system first rather slowly ebbs as the temperature rises, and then enhances. The temperature value corresponding to the minimum thermoelectric efficiency varies as the bridge length changes. The longer the molecular bridge becomes the smaller is this temperature. We remark that the fall of $ ZT $ accompanying the temperature rise is usually associated with  electron transport controlled by thermally assisted hopping of electrons between the bridge sites. Our results show that a similar fall of thermoelectric efficiency of a  single-molecule junction may occur within a certain temperature range even though the coherent tunneling remains the controlling transport mechanism.

\section{iii.  Conslusion}

In the present work we have analyzed the dependences of thermoelectric efficiency of a single-molecule junction on the length of the molecular linker. To compute the thermoelectric figure of  merit we simulated the bridge by a tight-binding chain. Obtained results show that $ ZT $ reveals a strong dependence on the bridge length. Also, it may be affected by gateway states which appear when characteristics of terminal sites on the bridge differ from those of remaining sites. Thermoelectric phenomena in unbiased metal-molecule junctions and other similar systems occur due to the difference in electrons distributions on the electrodes kept at different temperatures. Consequently, the profile of electron transmission near $ E = \mu $ takes on a very important part in controlling thermoelectric transport characteristics including $ ZT .$ As the molecular bridge lengthens, the transmission profile changes thus making thermoelectric characteristics dependent on the bridge length.

To thoroughly analyze thermoelectric properties of metal-molecule junctions more detailed models regarding the molecular electronic structure, coupling between metal electrodes and the molecular bridge and the effects of environment are needed. However, tight-binding models were successfully used to describe  length-dependent thermopower observed in some single-molecule junctions \cite{29,30,31}. This gives ground to believe that these simplified models capture some essential physics.   
                       Presently, there exists a significant interest in the interplay of spin effects and heat transport through quantum dots and metal-molecular junctions \cite{6,47,48}. It was demonstrated that spin effects may strongly affect Seebeck coefficient and ZT in systems with  ferromagnetic electrodes and/or those where the linker is a molecular magnet \cite{49,50,51,52}. Spin-crossover processes may also bring changes into the junction thermopower and cause ZT enhancement \cite{53}. The simplified approach developed in the present work using tight-binding models may be employed to explore length-dependent characteristics of spin-transport and spin-caloritronic effects in single-molecule junctions, bringing  further understanding of thermoelectric properties of nanoscale systems.  \vspace{2mm}

{\bf Acknowledgments:} The author thank G. M. Zimbovsky for help with the manuscript preparation. This work was supported by the US NSF-DMR-PREM 1523463.

\end{document}